\documentstyle[12pt]{article}
\def\ftoday{{\sl  \number\day \space\ifcase\month
\or Janvier\or F\'evrier\or Mars\or avril\or Mai
\or Juin\or Juillet\or Ao\^ut\or Septembre\or Octobre
\or Novembre \or D\'ecembre\fi
\space  \number\year}}
\newcommand{\journal}[4]{{\em #1~}#2\,(19#3)\,#4;}

\newcommand{\hpa}{\journal {Helv. Phys. Acta}}

\newcommand{\ijmp}{\journal {Int. J. Mod. Phys.}}

\newcommand{\pr}{\journal {Phys. Rev.}}

\newcommand{\cmp}{\journal {Comm. Math. Phys.}}

\newcommand{\np}{\journal {Nucl. Phys.}}
\newcommand{\pl}{\journal {Phys. Lett.}}

\makeatletter
\@addtoreset{equation}{section}
\makeatother
\renewcommand{\theequation}{\thesection.\arabic{equation}}
\newcommand{\dis}{\displaystyle}
\def\Lp{\displaystyle{\biggl(}}
\def\Rp{\displaystyle{\biggr)}}
\def\LP{\displaystyle{\Biggl(}}
\def\RP{\displaystyle{\Biggr)}}
\newcommand{\lp}{\left(}\newcommand{\rp}{\right)}
\newcommand{\lc}{\left[}\newcommand{\rc}{\right]}
\newcommand{\lac}{\left\{}\newcommand{\rac}{\right\}}
\newcommand{\G}{\Gamma}
\newcommand{\D}{\Delta}
\renewcommand{\a}{\alpha}
\renewcommand{\b}{\beta}
\renewcommand{\d}{\delta}
\newcommand{\e}{\varepsilon}
\newcommand{\eb}{\bar\varepsilon}
\newcommand{\f}{\phi}
\newcommand{\F}{\Phi}

\newcommand{\g}{\gamma}
\newcommand{\k}{\kappa}
\renewcommand{\l}{\lambda}
\newcommand{\lb}{\bar\lambda}

\newcommand{\m}{\mu}
\newcommand{\n}{\nu}

\newcommand{\p}{\psi}
\renewcommand{\pb}{\bar\psi}
\newcommand{\r}{\rho}
\newcommand{\s}{\sigma} \renewcommand{\S}{\Sigma}

\newcommand{\z}{\zeta}
\renewcommand{\AA}{{\cal A}}
\newcommand{\BB}{{\cal B}}

\newcommand{\BS}{{\cal B}_\Sigma}
\newcommand{\CC}{{\cal C}}

\newcommand{\FF}{{\cal F}}

\newcommand{\NN}{{\cal N}}

\newcommand{\SS}{{\cal S}}

\newcommand{\WW}{{\cal W}}

\newcommand{\complex}{{\kern .1em {\raise .47ex
\hbox {$\scriptscriptstyle |$}}
    \kern -.4em {\rm C}}}
\newcommand{\real}{{{\rm I} \kern -.19em {\rm R}}}
\newcommand{\rational}{{\kern .1em {\raise .47ex
\hbox{$\scripscriptstyle |$}}
    \kern -.35em {\rm Q}}}
\renewcommand{\natural}{{\vrule height 1.6ex width
.05em depth 0ex \kern -.35em {\rm N}}}

\newcommand{\cb}{{\bar c}}

\newcommand{\half}{\frac 1 2}
\newcommand{\pa}{\partial}
\newcommand{\pad}[2]{{\frac{\partial #1}{\partial #2}}}
\newcommand{\fud}[2]  {{\displaystyle{\frac{\delta #1}{\delta #2}}}}
\newcommand{\dfrac}[2]{{\displaystyle{\frac{#1}{#2}}}}

\newcommand{\sla}{\raise.15ex\hbox{$/$}\kern -.57em}

\newcommand{\twiddle}{\lower.9ex\rlap{$\kern -.1em\scriptstyle\sim$}}

\newcommand{\vf}{{\varphi}}
\newcommand{\pam}{{\partial_\mu}}

\newcommand{\equ}[1]{(\ref{#1})}
\newcommand{\eq}{\begin{equation}}
\newcommand{\eqn}[1]{\label{#1}\end{equation}}
\newcommand{\eea}{\end{eqnarray}}
\newcommand{\eqap}{\begin{eqnarray}}
\newcommand{\eqanp}[1]{\label{#1}\end{eqnarray}}
\newcommand{\eqa}{\eq\ba{rl}}
\newcommand{\eqan}[1]{\ea\label{#1}\end{equation}}
\newcommand{\ba}{\begin{array}}
\newcommand{\ea}{\end{array}}
\newcommand{\eqac}{\begin{equation}\begin{array}{rcl}}
\newcommand{\eqacn}[1]{\end{array}\label{#1}\end{equation}}

\renewcommand{\pad}[2]{{\displaystyle{\frac{\partial #1}{\partial #2}}}}

\def\intx{\displaystyle{\int d^4 \! x \, }}

\newcommand{\es}{\\[3mm]}

\newcommand{\fb}{\bar\f}

\newcommand{\Nb}{\pb^{*}{}}

\newcommand{\da}{{\dot{\a}}}
\newcommand{\db}{{\dot{\b}}}
\newcommand{\dg}{{\dot{\g}}}

\def\smab{\s^{\hspace{1mm}\m}_{\a\db}}
\def\esmab{\e^\a\smab\eb^{\db}}

\newcommand{\BSn}[1]{\BS^{(#1)}}

\def\B0modd{\makebox[28mm]{$\BSn{0}$-modulo-$d$}}

\newcommand{\dpad}[2]{{\displaystyle \frac{\pa #1}{\pa #2}}}
\newcommand{\adot}{{\dot\alpha}}

\newcommand{\dsum}[1]{\dis{\sum_{#1}}}

\newcommand{\dfud}[2]{{\displaystyle{\frac{\delta #1}{\delta #2}}}}

\begin{document}%\noindent
\newcommand{\pic}{$\spadesuit\spadesuit$}

\newcommand{\sic}{$\clubsuit\clubsuit$}

% COMMANDES POUR L'ARTICLE  wzbreak.tex:
%*********************************************************
%########################################################

\def\Lc{\displaystyle{\biggl[}}\def\Rc{\displaystyle{\biggr]}}
\def\LC{\displaystyle{\Biggl[}}\def\RC{\displaystyle{\Biggr]}}
\newcommand{\h}{\hbar}
\newcommand{\kb}{{\bar \kappa}}
\renewcommand{\lb}{{\bar\lambda}}
\newcommand{\ml}{{m^{(\l)}}}
\renewcommand{\mp}[1]{{m_{{#1}}^{(\p)}}}
\newcommand{\mpb}[1]{{m_{{#1}}^{(\pb)}}}
\newcommand{\mff}[1]{{m_{{#1}}^{2(\f\f)}}}
\newcommand{\mffb}[1]{{m_{{#1}}^{2(\f\fb)}}}
\newcommand{\mfbfb}[1]{{m_{{#1}}^{2(\fb\fb)}}}
\newcommand{\Ml}{{M^{(\l)}}}
\newcommand{\Mp}[1]{{M_{{#1}}^{(\p)}}}
\newcommand{\Mlb}{{M^{(\lb)}}}
\newcommand{\Mpb}[1]{{M_{{#1}}^{(\pb)}}}
\newcommand{\Mff}[1]{{M_{{#1}}^{(\f\f)}}}
\newcommand{\Mffb}[1]{{M_{{#1}}^{(\f\fb)}}}
\newcommand{\Mfbfb}[1]{{M_{{#1}}^{(\fb\fb)}}}

\newcommand{\na}{\nabla}
\newcommand{\dslav}{{\D_{\rm ST}}}
\newcommand{\zpara}[1]{{Z^{#1}}}
\newcommand{\para}[1]{{\rho_{#1}}}
\newcommand{\bpara}[1]{{{\bar \rho}_{#1}}}

\newcommand{\paras}[1]{{\rho^{\rm S}_{#1}}}
\newcommand{\bparas}[1]{{{\bar\rho{}^{\rm S}}_{#1}}}
\newcommand{\paraa}[1]{{\rho^{\rm A}_{#1}}}
\newcommand{\bparaa}[1]{{{\bar \rho{}^{\rm A}}_{#1}}}

\newcommand{\ul}{u}
\newcommand{\bul}{{\bar u}}
\newcommand{\vl}{v}
\newcommand{\bvl}{{\bar v}}
\newcommand{\vhat}{{\hat v}}\newcommand{\Vhat}{{\hat V}}
\newcommand{\vbhat}{{\hat{\bar v}}}\newcommand{\Vbhat}{{\hat{\bar V}}}
\newcommand{\ub}{{\bar u}} \newcommand{\vb}{{\bar v}}
\newcommand{\Ub}{{\bar U}} \newcommand{\Vb}{{\bar V}}
\newcommand{\?}{{\bf ???}}
\newcommand{\slav}{Slavnov-Taylor}
\newcommand{\susy}{supersymmetry}
\newcommand{\Susy}{Supersymmetry}
\newcommand{\susyc}{supersymmetric}
\newcommand{\Susyc}{Supersymmetric}
%########################################################
% UN  o  AU-DESSUS D'UNE LETTRE, COMME DANS `Angstrom':
\newcommand{\nmGA}{{\G\kern-.45em \raisebox{3.3mm}{\tiny o}\kern.30em}}
\newcommand{\nmSI}{{\S\kern -.50em \raisebox{3.3mm}{\tiny o}\kern.30em}}
\newcommand{\nmSS}{{\SS\kern -.50em \raisebox{3.3mm}{\tiny o}\kern.30em}}
\newcommand{\nmBB}{{\BB\kern -.50em \raisebox{3.3mm}{\tiny o}}}
\newcommand{\nmGAind}{{\G\kern-.37em\raisebox{2.1mm}{\tiny o}}}
\newcommand{\nmSIind}{{\S\kern-.37em\raisebox{2.1mm}{\tiny o}}} 
\newcommand{\nnmGAind}{{\kern.15em\nmGAind}} 
\newcommand{\nnmSIind}{{\kern.10em\nmSIind}}
\newcommand{\nmBnmG}{{\nmBB_\nnmGAind\kern.20em}} 
\newcommand{\nmBnmS}{{\nmBB_\nnmSIind\kern.20em}}  
\newcommand{\nmBS}{{\nmBB_{\kern.5mm\S}}}
\newcommand{\nmBG}{{\nmBB_{\kern.5mm\G}}} 
\newcommand{\nmBF}{{\nmBB_{\kern.5mmF}}} 
\newcommand{\nmfBS}[1]{\nmBB_\nnmSIind^{(#1)}}
\newcommand{\nmfBG}[1]{\nmBB_\nnmGAind^{(#1)}}
%########################################################
%########################################################
\titlepage  \noindent
hep-th/9604002\\
UGVA-DPT-1996-04-917 \vspace{20mm}

\begin{center}               
{\huge Algebraic Renormalization of $N=1$\\[4mm]
   Supersymmetric Gauge Theories with \\[4mm]
   Supersymmetry Breaking Masses\footnote{
Supported in part by the Swiss National Science Foundation.}}

\vspace{1cm}

{\Large Nicola Maggiore\footnote{On leave of absence from Universit\`a 
degli Studi di Genova, Dipartimento di Fisica.}, 
Olivier Piguet and Sylvain Wolf}
\vspace{1cm}

{\it D\'epartement de Physique Th\'eorique --
     Universit\'e de Gen\`eve\\24, quai E. Ansermet -- CH-1211 Gen\`eve
    4\\Switzerland}

\end{center}
\vspace{15mm}

\begin{center}
\bf ABSTRACT  
\end{center}  

{\it We provide $N=1$  Super Yang-Mills theory in the  Wess-Zumino gauge
with mass terms for the supersymmetric partners of the gauge fields 
and of the matter fields, together with a supersymmetric mass term
for the fermionic matter fields. All mass terms are chosen in such
a way to induce soft supersymmetry breakings at most, while preserving 
gauge invariance to all orders of perturbation theory. The breakings 
are controlled through  an extended  Slavnov-Taylor identity.  
The renormalization analysis, both in the ultraviolet and 
in the infrared region, is performed.
}
\vfill

%\newpage

%++++++++++++++++++++++++++++++++++++++++++++++++++++++++++++
\section{Introduction}
In a previous paper~\cite{I} we dealt
with the problem of renormalizing $N=1$ super Yang-Mills (SYM) in the
Wess-Zumino (WZ) gauge, to all orders of perturbation theory. We
restricted ourselves to the massless case with exact  supersymmetry. Our
main results were the expression of the most general counterterm  and of
the supersymmetric extension of the Adler-Bardeen-Bell-Jackiw gauge
anomaly. From the structure of the counterterm, on can infer a
nonvanishing $\b$-function for the physical parameters of the theory:
the gauge and the Yukawa couplings constants. Actually, it is known that
nonrenormalization properties hold
in some cases. In particular it has been
shown that a class of $N=1$ SYM theories satisfying certain criteria
have vanishing $\b$-function~\cite{lps}. 

It is of primary importance to investigate whether the mentioned
nonrenormalization properties are maintained in realistic supersymmetric
models, i.e. in theories in which the supersymmetry is broken by mass
terms for the superpartners of the presently known particles, namely for
the gauginos, the squarks and the sleptons. 
This program entails as a necessary
preliminary step the construction at the quantum level 
of such a massive  theory with broken
supersymmetry, which is the aim of the
present work.  As
explained in~\cite{I}, the main benefits of superspace are lost if
supersymmetry is broken. We therefore prefer to work in the WZ gauge,
which has the advantage of depending on a minimal number of fields and
which also turns out to be the preferred choice in the
phenomenological applications. 

The masses  being physical quantities,
the corresponding mass terms in the action must be chosen gauge
invariant. On the other hand some of them will necessarily 
break supersymmetry. Our task is to control the supersymmetry 
breakings
while preserving gauge invariance to all orders of perturbation theory,
by means of a renormalizable quantum field theory.

As in the massless case, the whole information is contained in a
generalized \slav\ operator, from which the symmetries of the
theory, such as BRS and \susy\ for instance,  can be recovered by making
a filtration in some global ghosts (see the next section or~\cite{I} for 
details). The masses which break \susy, 
are introduced through constant shifts of suitable external fields,
following a  procedure~\cite{blasi-magg}
which allows us to control easily the \susy\
breakings via a \slav\ identity, with quantum gauge invariance -- 
expressed by the usual BRS invariance -- guaranteed by a simple
functional identity.   

 We restrict ourselves to the
mass terms selected by Girardello and 
Grisaru (GG)~\cite{gir-gri} as yielding
``soft'' \susy\ breakings, i.e not inducing
ultraviolet (UV) divergences worse than logarithmic.

 An important property of $N=1$ \susyc\ theories is their invariance
under a phase transformation, not commuting with the \susy\ generators,
called  \mbox{$R$-invariance~\cite{r-invar}}.
It is at the basis of the supercurrent multiplet~\cite{fer-zum} 
which plays a crucial role in the proof of the nonrenormalization
of the coupling constants of some \susy\ gauge models~\cite{lps}.
 However $R$-invariance is broken by the mass terms. But this breakdown 
is also controlled by the \slav\ identity which takes into account 
the transformation properties under $R$ of
the same external fields already introduced 
for controlling the \susy\ breakings.
  Moreover, we also consider \susyc\ masses in order to keep  in with
the GG conditions, which would be violated by a supersymmetry breaking
matter fermion mass. Since they break $R$-invariance
as well, these masses are also introduced via similar external
fields. 

 Besides the  gauginos, all matter fields are made massive. 
Although this is not necessary it helps in avoiding 
infrared problems. Only the gauge vector
field and the Faddeev-Popov ghosts remain massless.

Gauge field and fermion masses produced by a Higgs mechanism are
not considered in the present paper. Gauge invariance is assumed exact,
without spontaneous breakdown.

 Due to the presence both of massless and 
of massive fields the proof of
the absence of infrared (IR) singularities needs a careful discussion.
This will be done in an appendix in order not to charge the main text
with technicalities. As we shall see it turns out that the construction
 based only on the ultraviolet (UV) power 
counting, yields a model free of IR singularity.

 The paper is organized as follows. In Section~\ref{resume} we give a
brief summary of~\cite{I}. In Section~\ref{brisures-de-masse} we  present
our method for introducing masses and write the general classical action
for massive $N=1$ SYM theory. Section~\ref{renormalisation} is
devoted to the quantum extension of the model, by proving its
renormalizability. The Callan-Symanzik equation is derived in 
Section~\ref{eq-de-callan-symanzik}. Finally we give an outlook in the
concluding Section~\ref{conclusion}. 
The absence of IR singularities is shown in  the Appendix. 

%##################################################################
\section{The Massless Case}\label{resume}
\subsection{The Massless Action}
The massless theory~\cite{I} is described by an effective action -- or vertex
functional -- 
\eq
\nmGA(A^i_\m,\l^i_\a,\f_a,\p_{a\a},c^i,\cb^i,b^i; 
        A^{*i}_\m,\l^{*i}_\a,\f^*_a,\p^*_{a\a},c^{*i};
        \xi^\m,\e_\a,\eta) =  \nmSI + O(\h)\ ,
\eqn{massless-gamma}
  where the superscript  on $\G$ and $\S$ means ``massless'', and
  $\nmSI$ is the corresponding classical action, equivalent to
the effective action at the tree level:
\eqa
\nmSI = \intx \LP &\!\!\!
    \displaystyle{\frac{1}{g^2}}\lp
    -\frac{1}{4}F^{i\m\n}F^i_{\m\n}
    -i\l^{i\a}\smab(D_\m\lb^\db)^i \rp
    -\frac{1}{8}g^2 |\fb_aT^i_{ab}\f_b|^2
    +\frac{1}{2}|D_\m\f|^2
\nonumber\es
&\!\!\!
  -i\p^\a_a\smab(D_\m\pb^\db)_a
  -2\l_{(abc)}\f_b\f_c\lb_{(ade)}\fb_d\fb_e
  -i\pb_{a\db}T^i_{ab}\f_b\lb^{i\db }
\nonumber\es
&\!\!\!
  -i\l^{i\a}\fb_aT^i_{ab}\p_{b\a}
  +2\l_{(abc)}\p^\a_a\p_{b\a}\f_c
  +2\lb_{(abc)}\pb_{a\db}\pb^\db_b\fb_c 
     \nonumber
\nonumber\es
&\!\!\!
+  b^i\partial^\m A^i_\m +  (\partial^\m\cb^i)\lp(D_\m c)^i
  +\e^\a\s_{\m\a\db}\lb^{i\db} +\l^{i\a}\s_{\m\a\db}\eb^\db\rp
\nonumber\es
&\!\!\!
+ A^{*}{}^{i\m}(sA^i_\m) + \l^{*}{}^{i\a}(s\l^{i}_{\a})
+ \lb^{*}{}^i_\db(s\lb^{i\db})
+ \f^{*}_a(s\f_a) + \fb^{*}_a(s\fb_a) + \p^{*}{}^{\a}_{a}(s\p_{a\a})
\nonumber\es &\!\!\!
+ \Nb_{a\db}(s\pb^\db_a)
+c^{*i}(sc^i) -\frac{g^2}{2}(\e^\a \l^{*}{}^i_\a -\eb_\db\lb^{*}{}^{i\db})^2
+2\e^\a \p^{*}{}_{a\a}\eb_\db\Nb^\db_a \RP\ .
\eqan{class-action-massless}
 Here $s$ is an extended BRS operator which puts together 
the gauge, the \susy\ and $R$-transformations
as well as the translations. We
refer the reader to~\cite{I} for the expression of the $s$-variation of the
various fields and global ghosts. The latter are the following:
\begin{enumerate} 
\item[-] the gauge and gaugino fields $A$ and $\l$, 
in the adjoint representation of the gauge group,
with running index $i$,
\item[-] the scalar and fermion matter fields $\f$ and $\p$, in some
(reducible) representation, with running index $a$,
\item[-] the ghost, antighost and Lagrange multiplier
fields $c$, $\cb$ and $b$, in the adjoint representation,
\item[-] the ``antifields'' $A^*$, $\l^*$, $\f^*$, $\p^*$  and $c^*$,
i.e. the external fields coupled to the BRS transformations 
of the fields which transform nonlinearly under BRS,
\item[-] the coordinate independent infinitesimal
parameters $\xi$, $\e$ and $\eta$ of the translations, 
\susy\ transformations and
$R$-transformations, respectively, promoted to the rank of
``global ghosts'', their Grassmann parity being chosen as
opposite to the natural one.
\end{enumerate}
The spinor fields are in the Weyl representation, with spinor index $\a$
= 1,2 (and dotted index $\da$ = 1,2 for their conjugates).

The dimensions, Grassmann parities, ghost numbers  and $R$-weights of
all the fields and global ghosts
are shown in Table~\ref{table-dim}. 
More details on our
conventions and notations can be found in  Appendix C of~\cite{I}.
\begin{table}[hbt]
\centering
\begin{tabular}{|c||c|c|c|c|c|c|c|c|c|c|c|c|c|c|c|}
\hline
&$A_\mu$&$\lambda$&$\f$&$\p$&$c$&$\cb$&$b$&$\xi^\m$
&$\e$&$\eta$&$A^*_\m$&$\l^*$&$\f^*$&$\p^*$ &$c^*$ \\
\hline\hline
$d$&1&3/2&1&3/2&0&2&2&-1&-1/2&0&3&5/2&3&5/2&4 \\
\hline
$r$&1&2&2&2&0&2&2&-1&-1/2&0&3&7/2&3&5/2&4 \\
\hline
$GP$&0&1&0&1&1&1&0&1&0&1&1&0&1&0&0 \\
\hline
$\F\Pi$&0&0&0&0&1&-1&0&1&1&1&-1&-1&-1&-1&-2\\
\hline
$R$&0&-1&-2/3&1/3&0&0&0&0&-1&0&0&1&2/3&-1/3&0 \\
\hline
\end{tabular}
\caption[t1]\hfill{\small\parbox{15cm}{ \ \\
 UV dimensions  $d$, IR dimensions $r$,
Grassmann parity $GP$, ghost numbers $\F\Pi$
  and \mbox{R-weights.} The IR dimensions will be defined and
used in the Appendix. In the main text the word ``dimension'' refers to
the UV dimension.}}
\label{table-dim}
\end{table}

The extended BRS operator $s$ in \equ{class-action-massless} is
nilpotent only on-shell, but the terms  in the classical action which are
quadratic in the external fields $\l^*$  and $\p^*$ 
allow the off-shell extension 
of the latter property. The action~\equ{class-action-massless} 
is indeed the most general
solution of a \slav\ identity, which has been
shown in~\cite{I} to hold for the effective action $\nmGA$ to all orders
of perturbation theory:
\eq
\nmSS(\nmGA) = 0\ ,
\eqn{slavnov-massless}
where the \slav\ operator of the massless theory is defined,  for
any functional $F$, by
\eqa
\nmSS(F):=&\nonumber\\[3mm]
\intx\LP &\!\!\!\!
\fud{F}{A^{*}{}^{i\m}}
\fud{F}{A^i_\m}
+
\fud{F}{\l^{*}{}^i_\a}
\fud{F}{\l^{i\a}}
+
\fud{F}{\lb^{*}{}^{i\db}}
\fud{F}{\lb^{i}_{\db}}
+
\fud{F}{\f^{*}_a}
\fud{F}{\f_a}
+
\fud{F}{\fb^{*}_a}
\fud{F}{\fb_a}
+
\fud{F}{\p^{*}{}_{a \a}}
\fud{F}{\p^\a_a}
\nonumber\\[3mm]&\!\!\!\!\!\!\!
+
\fud{F}{\Nb_{a}^{\db}}
\fud{F}{\pb_{a\db}}
+
\fud{F}{c^{*i}}
\fud{F}{c^i}
+
(b^i-i\xi^\m\pa_\m\cb^i)\fud{F}{\cb^i}
\nonumber\\[3mm]&\!\!\!\!\!\!\!
+
\lp -2i\esmab\pa_\m\cb^i
-i\xi^\m\pa_\m b^i\rp\fud{F}{b^i} \RP
-2\esmab\pad{F}{\xi^\m}
-\eta\e^\a\pad{F}{\e^\a}
-\eta\eb_\db\pad{F}{\eb_\db}\ .
\eqan{slav-op-massless}
The corresponding ``linearized'' \slav\ operator reads
\eqa
\nmBF:=&\nonumber\\[3mm]
\intx\LP &\!\!\!\!
\fud{F}{A^{*}{}^{i\m}}
\fud{}{A^i_\m}
+
\fud{F}{A^i_\m}
\fud{}{A^{*}{}^{i\m}}
+
\fud{F}{\l^{*}{}^i_\a}
\fud{}{\l^{i\a}}
-
\fud{F}{\l^i_\a}
\fud{}{\l^{*}{}^{i\a}}
+
\fud{F}{\lb^{*}{}^{i\db}}
\fud{}{\lb^{i}_{\db}}
-
\fud{F}{\lb^{i\db}}
\fud{}{\lb^{*}{}^{i}_{\db}}
\nonumber\\[3mm] &\!\!\!\!
+
\fud{F}{\f^{*}_a}
\fud{}{\f_a}
+
\fud{F}{\f_a}
\fud{}{\f^{*}_a}
+
\fud{F}{\fb^{*}_a}
\fud{}{\fb_a}
+
\fud{F}{\fb_a}
\fud{}{\fb^{*}_a}
+
\fud{F}{\p^{*}{}_{a \a}}
\fud{}{\p^\a_a}
-
\fud{F}{\p_{a\a}}
\fud{}{\p^{*}{}^{\a}_a}
\nonumber\\[3mm] &\!\!\!\!
+
\fud{F}{\Nb_{a}^{\db}}
\fud{}{\pb_{a\db}}
-
\fud{F}{\pb_{a}^{\db}}
\fud{}{\Nb_{a\db}}
+
\fud{F}{c^{*i}}
\fud{}{c^i}
+
\fud{F}{c^i}
\fud{}{c^{*i}}
+
(b^i-i\xi^\m\pa_\m\cb^i)\fud{}{\cb^i}
\\[3mm]&\!\!\!\!
+
(-2i\esmab\pa_\m\cb^i -i\xi^\m\pa_\m b^i)\fud{}{b^i} \RP
-2\esmab\pad{}{\xi^\m}
-\eta\e^\a\pad{}{\e^\a}
-\eta\eb_\db\pad{}{\eb_\db}\ .\nonumber
\eqan{slavlin-massless}
The fulfillment of the \slav\ identity \equ{slavnov-massless} implies the
nilpotency property
\eq
(\nmBnmG)^2= 0\ .
\eqn{nilp-massless}

By filtrating the linearized \slav\ operator with the operator
\eq
\NN = \e^\a\pad{}{\e^\a} + {\bar\e}^\adot\pad{}{{\bar\e}^\adot}
   + \xi^\m\pad{}{\xi^\m}
  + \eta\pad{}{\eta} \ ,
\eqn{a-filter}
which counts the number of global ghosts, we can extract each quantum
symmetry by means of an algebraic identification. Writing
\eq
\nmBnmG =\dsum{n\ge0}\nmfBG{n}\ ,
    \quad{\rm with}\quad \lc \NN,\nmfBG{n}\rc= n \nmfBG{n} \ ,
\eqn{a-filtr-bb}
we identify $\nmfBG{0}$ as the usual 
quantized gauge BRS operator, whereas
$\nmfBG{1}$ contains the functional operators 
(``Ward identity
operators'') $\WW_\a$, $\WW_\m$ and $\WW_R$
of supersymmetry, translation and $R$-invariance,
respectively: 
\eq
\nmfBG{1} = 
  \e^\a \WW_\a + {\bar\e}^\adot {\bar\WW}_\adot + \xi^\m \WW_\m
  + \eta\WW_R
  -2 \e^\a\s^\m_{\a\adot}{\bar\e}^\adot\pad{}{\xi^\m}
  -  \eta\e^\a\pad{}{\e^\a}
  +  \eta{\bar\e}^\adot\pad{}{{\bar\e}^\adot}\ .
\eqn{a-def-symm-gen}
This means that, at any order of perturbation theory, we can skip from
the language represented by the generalized \slav\ operator
\equ{slav-op-massless} to the description in terms of the single Ward
operators $\WW$.  Let us recall that the 
algebra obeyed by the latter is not closed 
(see Section 4 of~\cite{I}).

%#########################################################
%\section{Gauge Fixing and All That}
\subsection{Gauge Condition, Ghost Equation and Global Ghost 
Equations}\label{fixation-de-jauge,-etc.}
  Since the identities written in this and  in the following subsection are
not altered by the introduction of the masses we drop here the
superscript ${}^{0}$ on $\G$ and $\S$.

The gauge fixing condition is of the Landau type:
\eq
\fud{\S}{b^i} = \pa^\m A^i_\m \ .
\eqn{gaugecond}
The ghost equation, peculiar to the Landau gauge, reads
\eq
\FF^i\S := \intx\lp\dfud{}{c^i}-f^{ijk}\cb^j\dfud{}{b^k}\rp \S 
   = \D^i_{\rm g} \ ,
\eqn{ghostexpr}
where
\eqa
\D^i_{\rm g} := \intx\LP &\!\!\!
f^{ijk}\Lp -A^{*}{}^{j\m}A^k_\m +\l^{*}{}^{j\a}\l^k_\a
+\lb^{*}{}^j_\db\lb^{k\db} +c^{*j}c^k \Rp
\nonumber\es &\!\!\!
+T^i_{ab} \Lp \f^{*}_a\f_b +\fb^{*}_a\fb_b
-\p^{*}{}^\a_ a\p_{b\a} -\Nb_{a\db}\pb^\db_b
\Rp\RP
\eqan{ghostbr}
is a classical breaking, i.e. is linear in the dynamical fields.

The global ghost equations are
\eq
\pad{\S}{\xi^\m} = \D^{\rm t}_\m\ \ ,\ \ \pad{\S}{\eta} = \D_{\rm R}\ ,
\eqn{globalgh} 
where $\D^{\rm t}_\m$ and $\D_{\rm R}$ are classical breakings given by
\eqa
\D^{\rm t}_\m := -i\,\intx \LP&-A^{*}{}^i_\n\pam A^{i\n}
+\l^{*}{}^{i\a}\pam \l^i_\a
+\lb^{*}{}^i_\db\pam \lb^{i\db}+c^{*i}\pam c^i\nonumber\es
&-\f^{*}_a\pam\f_a-\fb^{*}_a\pam\fb_a+\p^{*}{}^\a_a\pam\p_{a\a}
+\Nb_{a\db}\pam\pb^\db_a\RP ,
\eqan{globalbr1}
\eq
\D_{\rm R} := \intx \lp -\l^{*}{}^{i\a}\l^i_\a +\lb^{*}{}^i_\db\lb^{i\db}
+\frac{2}{3}\f^{*}_a\f_a-\frac{2}{3}\fb^{*}_a\fb_a
+\frac{1}{3} \p^{*}{}^\a_a\p_{a\a}-\frac{1}{3}\Nb_{a\db}\pb^\db_a
\rp .
\eqn{globalbr2}
The two global ghost equations express the linearity of the translations
and of the $R$-transformations, respectively.

%######################################################
\subsection{Algebra, Antighost Equation and Rigid Invariance}
The \slav\ operator $\SS$, the functional derivative 
$\d/\d b$, the ghost operator $\FF^i$ and the partial derivatives 
$\pa/\pa\xi^\m$,  $\pa/\pa\eta$ obey an algebra together with the 
following operators:

\noindent 
The antighost operator
\eq
{\bar{\FF}}^i := \fud{}{\cb^i} + \pam \fud{}{A^{*}{}^i_\m} +
                  i\xi^\m\pa_\m \fud{}{b^i}\ ,
\eqn{antighost}
the Ward operator for the rigid transformations:
\eq
\WW^i_{\rm rig} := \intx  \lp\dsum{\vf}
\d^i_{\rm rig}\vf \dfud{}{\vf}\rp \ ,
\eqn{wrig}
the translation Ward operator:
\eq
\WW_\m := -i \intx \lp\dsum{\vf}
\pam\vf\dfud{}{\vf}\rp  
\eqn{wmu}
and the Ward operator for the $R$-transformations:
\eq
\WW_{\rm R} := \intx  \lp\dsum{\vf}
R_\vf \vf\dfud{}{\vf}\rp \ .
\eqn{wr}
The summation over $\vf$ in \equ{wrig} includes 
all the fields listed in Table~\ref{table-dim}. Here 
$\d^i_{\rm rig}\vf$ is the infinitesimal rigid transformation of $\vf$, the
gauge, gauginos and (anti)ghost fields transforming in the adjoint 
representation of the
gauge group and the matter fields in their own representation.
In \equ{wmu} and \equ{wr} the summation also includes
the external doublets $(u,v)$, $(U,V)$ 
  to be introduced in the next section. $R_\vf$
is the $R$-weight of the field $\vf$.

The algebra reads
%*******0*****
\eq
\BB_F \SS(F)  = 0 ,
\eqn{alg0}
%*******1*****
\eq
\fud{}{b^i}\SS(F)- \BB_F\lp\fud{F}{b^i}-\pam A^{i\m}\rp=\bar{\FF^i} F  ,
\end{equation}
%*******2*****
\eq
{\bar{\FF}}^i\SS(F) +  \BB_F {\bar{\FF}}^i F = 0 ,
\end{equation}
%*******3*****
\eq
\FF^i \SS(F)+\BB_F\lp\FF^i F - \D^i_{\rm g}\rp = \WW^i_{rig}F ,
\end{equation}
%*******4*****
\eq
\WW^i_{rig}\SS(F) - \BB_F\WW^i_{rig}F = 0 ,
\end{equation}
%*******5*****
\eq
\pad{}{\xi^\m} \SS(F)+ \BB_F \lp\pad{F}{\xi^\m}-\D^{\rm t}_\m\rp=\WW_\m F
,
\end{equation}
%*******6*****
\eq
\WW_\m \SS(F) - \BB_F \WW_\m F = 0 ,
\end{equation}
%*******7*****
\eq
\pad{}{\eta} \SS(F)+\BB_F \lp\pad{F}{\eta}-\D_{\rm R}\rp = \WW_{\rm R} F ,
\end{equation}
%*******8*****
\eq
\WW_{\rm R} \SS(F) -  \BB_F \WW_{\rm R} F = 0  ,
\eqn{alg8}
%*************
where $F$ is an arbitrary functional.

Setting $F=\S$ and using the \slav\ identity and the identities
\equ{gaugecond}, \equ{ghostexpr} and \equ{globalgh} satisfied by the
classical action $\S$, we get the antighost equation and the Ward
identities for the linear symmetries, i.e. for the rigid, translation
and $R$-symmetries, respectively:
\eq
{\bar{\FF}}^i \S = 0 , \quad
\WW^i_{\rm rig} \S = 0 , \quad
\WW_\m  \S = 0 , \quad
\WW_{\rm R} \S = 0 .
\eqn{antigh-and-WI}

%##############################################################
\section{Massive $N=1$ SYM Theory}\label{brisures-de-masse}
\subsection{Generalities}\label{introd.-des-masses}
  Let us begin with a general description of the procedure we
shall follow for introducing a generic mass term. 

Masses are introduced, in the classical theory,
in two ways: the first one breaks \susy,
the second one does not. The aim of the former is to give masses to the
gauginos and to the scalar matter fields. 
These breakings are known as soft in the
classification of~\cite{gir-gri}. The aim of the latter is to give
masses to the fermion matter fields 
without introducing \susy\ breakings which would not
be ``soft''. 
Only the gauge fields remain
massless.  Let us remark that  all these masses break 
$R$-symmetry.  However, the addition of the mass terms to the massless 
action \equ{class-action-massless} has to 
preserve gauge invariance,  also at the quantum level.
%++++++++++++++++++++++++++++++++++++++++++++++++++++++++++++
\subsubsection{\Susy\ Breaking Masses}\label{masses-brisant-susy} 

  A generic \susy\ breaking mass term is denoted by 
\eq
\S_m = \intx M(x)\ ,
\eqn{gen-mass-term} 
where $M(x)$ is some quadratic expression in the matter fields or in the
gaugino field.
Its gauge invariance is expressed by  the
invariance under the gauge BRS operator $\nmfBS{0}$ 
defined by the zeroth order term of the filtration \equ{a-filtr-bb} --
with $\nmGA$ replaced by the massless action
$\nmSI$ since we are dealing here with the
classical theory. Thus
\eq
\nmfBS{0} \S_m = 0 \ .
\eqn{inv-mass-term}
This can equivalently be written as 
\eq
\nmBnmS \S_m = O(\e,\eta)\ ,
\eqn{inv-mass-term'}
the terms in $\e$ of the right-side being due to the breakdown of
\susy\ and those in $\eta$ to the breakdown of
$R$-invariance. 

Adding the mass terms \equ{gen-mass-term} to the massless action $\nmSI$, 
we demand that the new action $\S$
obeys a new, exact, \slav\ identity:
\eq
\SS(\S) = 0\ .
\eqn{class-slavnov}
 The idea is to couple each mass term \equ{gen-mass-term} to a
doublet $(u,v)$ in a BRS invariant way, i.e. in a such
a way that the new \slav\ identity \equ{class-slavnov} holds, with $\SS$
defined by
\eq
\SS(F) = \nmSS(F) +  \dsum{u,v} \intx
   \lp s u \dfud{F}{u} + s v\dfud{F}{v} \rp\ , 
\eqn{ext-slavnov-op}
the summation being performed over all doublets $(u,v)$ -- 
and their complex conjugates -- associated to each
mass term. The Grassmann
parities of $u$, resp. $v$ are negative, resp. positive.
The nilpotent BRS
variations $su$ and $sv$ are given by
\eq\ba{l}
s u(x) = v(x)+\k_u-i\xi^\m\pa_\m u(x) + \eta R_u u(x) \ ,
    \quad s\bar u = -\overline{su}\ ,\es
s v(x) = -2i\e\s^\m\eb\pa_\m u(x) -i\xi^\m\pa_\m v(x) + 
           \eta R_u (v(x)+\k_u)\ ,
           \quad s\bar v = \overline{sv} \ .
\ea\eqn{BRS-u-v}
The external fields $v$ are shifted by the dimensionful constants $\k_u$ 
which parameterize the masses. The numbers $R_u$ are the
$R$-weights of the doublets $(u,v)$, chosen opposite to those of the
mass terms \equ{gen-mass-term} in order to keep the formal $R$-invariance.

The associated linearized \slav\ operator is
\eq
\BB_F = \nmBF + \dsum{u,v} \intx
   \lp s u \dfud{}{u} + s v\dfud{}{v} \rp\ ,
\eqn{lin-slavnov-op}
$\nmBF$ being defined by \equ{slavlin-massless}. Again the validity of
the \slav\ identity \equ{class-slavnov} implies the nilpotency condition
\eq
(\BB_\S)^2= 0\ .
\eqn{BS-nilpotency}
The identity
\equ{class-slavnov} taken at $u=v=0$ reads
\eq
\nmSS(\S)\Big\vert_{u=v=0} = -\dsum{u,v}\,\k_u\left.\intx
  \lp\dfud{\S}{u} + \eta R_u\dfud{\S}{v}\rp \right\vert_{u=v=0}\ .
\eqn{class-brok-slavnov}
This is the \slav\ identity  explicitly
broken by the mass terms. The first term in the right-hand side
represents the \susy\ breaking, whereas the second one represents the
breaking of $R$-invariance.

As we said above we want to keep the usual
gauge invariance. We therefore impose the supplementary 
condition
\eq
X\S^{(0)} := \left.\intx\dfud{\S}{u}\right\vert_{\e=\xi=\eta=0}=0\ ,
\eqn{gauge-inv-mass}
which ensures the validity of the gauge \slav\ identity
\eq
\left.\nmSS(\S)\right\vert_{\e=\xi=\eta=0}=0\ .
\eqn{class-gauge-inv}
The general solution of the full \slav\ identity  \equ{class-slavnov} -- 
the gauge fixing condition, ghost equation, etc. being taken into
account -- has the structure
\eq\ba{l}
\S = \nmSI + \dsum{u,v} \lp \BB_\nmSIind\intx uM_u(x) + \D_u \rp \es
\phantom{\S} = \nmSI + 
 \dsum{u,v}\lp \intx\lp (v+\k_u)M_u(x) -
                u\BB_\nmSIind M_u(x)\rp + \D_u \rp\ ,
\ea\eqn{struct-gen-sol}
where the $M_u(x)$ are
``mass terms'' such as the one introduced in \equ{gen-mass-term},
gauge invariant due to the condition \equ{gauge-inv-mass}.
The $\D_u$ are corrections needed due to the 
nonlinearity of the operator $\SS$.  

%##########################################################
\subsubsection{\Susyc\ Masses}\label{masses-susy}
Mass terms 
\eq
\S_{m,\,\rm SUSY} = \intx M_{\rm SUSY}(x)\ ,
\eqn{susy-mass-term} 
which do not break \susy, are also considered. They are
introduced in the same way as the non\susyc\ ones, i.e. through 
a doublet of external fields, now denoted by
$(U,V)$, with the same transformation laws \equ{BRS-u-v} as the 
generic doublet $(u,v)$. The \slav\ operator
\equ{ext-slavnov-op}, as well as its linearized form
\equ{lin-slavnov-op}, are completed accordingly. The $V$ fields are
shifted, too, by amounts $\k_U$ which parameterize the
supersymmetric masses. 

We need a condition ensuring the invariance under \susy\ of this
kind of 
mass terms. Taking into account the fact that the \susy\ invariance of 
a mass term such as \equ{susy-mass-term} is expressed by the fact that
its integrand $M_{\rm SUSY}(x)$ 
transforms as a total derivative under the \susy\ transformations, we 
can write this condition as
\eq
Y\S := \intx\dfud{\S}{U} = 0\ .
\eqn{susy-cond} 
Indeed, the \slav\ identity \equ{class-brok-slavnov}, taken
at $U=V=0$ shows only the 
$R$-breaking term in its right-hand side. Moreover a gauge
invariance condition such as \equ{gauge-inv-mass} is not needed 
 in this case since it 
is already implied by the condition \equ{susy-cond}.

%#############################################################
\subsection{The Massive Action}
Two types of masses are introduced:
\begin{enumerate}
\item[i)]{\bf \Susy\ breaking masses belonging to the GG class:}
i.e. masses which do not generate 
UV divergences worse than logarithmic. Their exhaustive list, given
in \cite{gir-gri}, consists of a gaugino mass term
and of two scalar field mass terms:
\eq\ba{l}
\Ml =  \l^{i\a}\l^i_\a   \ ,\es
\Mffb{ab} =  \f_a\fb_b  \ ,\es
\Mff{ab} = \f_a\f_b  \ .
\ea\eqn{GG-mass-terms} 
\item[ii)]{\bf \Susyc\ masses:}
i.e. the \susy\ invariant terms 
\eq
M^{\rm SUSY}_{ab} = 
    \p_a^\a\p_{b\a} - \lb_{(acd)}\f_b\fb_c\fb_d 
                    - \lb_{(bcd)}\f_a\fb_c\fb_d
   + \eb_\da\pb^{*\da}_a\f_b + \eb_\da\pb^{*\da}_b\f_a \ ,
\eqn{susy-mass}
\end{enumerate}
In order to control the breakings of supersymmetry and $R$-invariance
induced by the mass terms, the latter are coupled to doublets
of shifted external fields  
\eq%\ba{l} 
(u \ ,\vhat:=v+\k)\ ,\quad
(U_{(ab)}\ ,\Vhat_{(ab)}:=V_{(ab)}+K_{(ab)})\ ,
\eqn{ext-fields-doublets}
following the scheme defined in Subsection \ref{introd.-des-masses}. 
Because of the gauge invariance of the mass terms,
$u$ and $v$ are 
singlets of the gauge group, whereas $U_{ab}$ and $V_{ab}$ are
 symmetric invariant tensors.
 The shifts $K_{ab}$ are in
general nondiagonal if some matter supermultiplets  
belong
to the same irreducible representation of the gauge group.
The dimensions, \mbox{$R$-weights}, Grassmann parities and
ghost numbers are shown in Table~\ref{table-dim-u-v}.  
%---------------------------------------------------------------------
\begin{table}[hbt]
\centering
\begin{tabular}{|c||c|c|}
\hline
&$u,v$&$U_{ab},V_{ab}$ \\
\hline\hline
$d=r$&1        &1 \\
\hline
$R$&2          &$-2/3$ \\
\hline
\end{tabular}
\caption[t2]\hfill{\small\parbox{15cm}{ \ \\
 Ultraviolet dimensions $d$, 
 infrared dimensions $r$ (see the Appendix) 
and \mbox{R-weights} of the doublets of
complex external fields. 
The Grassmann parity of the fields $u$ and $U_{ab}$ is odd, 
that of the fields  $v$ and $V_{ab}$ is even. 
The formers have ghost number $-1$, 
the latter have ghost number 0.}}
\label{table-dim-u-v}
\end{table}
%-----------------------------------------------------------
The doublet $(u,\vhat)$ is coupled to the gaugino mass term 
$\Ml$ and the doublet 
$(U_{ab},\Vhat_{ab})$ to the supersymmetric
mass term $M^{\rm SUSY}_{ab}$. 
We did not introduce a doublet of external fields
for each of the mass terms listed
above. The reason, as we shall see, is that
the mass terms $\Mffb{ab}$ and $\Mff{ab}$ will
appear coupled to the doublets already
introduced.
%##############################################################

  The complete classical action defined accordingly 
to the scheme defined by \equ{struct-gen-sol}, taking into
account the possible mixing
compatible with power counting and $R$-inv\-ar\-iance, reads
\eq\ba{l}
\S  = \nmSI + \BB_\nmSIind\intx\Lp 
      \ul \lp \Ml + \para{1}{}_{(abc)}\f_a\f_b\f_c
              +\para{2}{}_{ab} \f_a\e^\a\p^*{}_{b\a} 
              + \para{3}{}_{ab} \vbhat \Mffb{ab} \right.\es
\phantom{\S  = \nmSI}\left.
     +\para{4}{}_{ab}\ub\e^\a \p_{a\a}\fb_b
     +\para{5}{}_{(ab)(cd)}\Vhat_{ab}\Mff{cd}
     +\para{7}\e^\a\s^\m_{\a\adot}\lb^{*i\adot}A^i_\m \rp\es
\phantom{\S  = \nmSI}                  
     + U_{ab} \lp
      M^{\rm SUSY}_{ab} 
     -\para{6}{}_{(ab)c}^i\pa^\m (\fb_c A^i_\m) \rp
     \ \Rp \ +\ \mbox{c.c.} \ 
     + \D  \ ,
\ea\eqn{complete-action}
%\ea\eqn{GG-complete-action}
 where the coefficients $\para{}$ have symmetry properties of their
indices which are explicitly shown by the parentheses. Moreover the
matrix $\para{3}$  can be chosen hermitian: 
$\bpara{3}{}_{ab}=\para{3}{}_{ba}$.  Indeed, one can see that
its antihermitian part may be absorbed in $\para{4}$.
 Besides the \slav\ invariance, we have also imposed on 
$\S$ the gauge invariance 
condition \equ{gauge-inv-mass} as 
well as the \susy\ condition \equ{susy-cond} for $U=U_{ab}$.
The last term $\D $, whose presence is due to the nonlinearity of the
\slav\ identity, reads
\eq\ba{l}
\D  = \intx\LP 4g^2\ul\bul \e^\a\l^i_\a\eb_\db\lb^{i\db} \es
\phantom{\D} 
-\dfrac{1}{2}\lp\para{2}{}_{ba}\lp\vhat\f_b-2u\e^\a\p_{b\a}\rp
                  -2\Vbhat_{ab}\fb_b\rp 
  \lp\bpara{2}{}_{ca}\lp\vbhat\fb_c-2\bar u\eb_\da\pb^\da_c\rp
     -2\Vhat_{ac}\f_c\rp \es \phantom{\D}
+\Lc 2\para{7}\bpara{7}\ul\bul\esmab A^i_\m \e^\b\l^*{}^i_\b
+\frac{i}{g^2}\para{7}\bpara{7}  
 \e^\a\smab {\bar \s}^\n {}^{\db\g} \s^\r_{\g\dg} \eb^\dg
   \ul D_\n (\bul A^i_\r ) A^i_\m   \es \phantom{\D}
+2\bpara{7} \ul \l^{i\a}\smab \eb^\db\lp
 \bvl A^i_\m + \bul (\pam c^i + \e^\g \s_\m {}_{\g\dg} \lb^{i\dg} ) \rp
 -i\bpara{7} \bul \esmab A^i_\m \para{2}{}_{ca} T^i_{ab} \ul \f_c\fb_b
  \es \phantom{\D}
-i\lp\para{6}{}_{(ab)c}^i\pam U_{ab}\fb_c
   +\bpara{6}{}_{(ab)c}^i\pam {\bar U}_{ab}\f_c\rp
  \para{7}\ul \e^\a\s^{\m\n}{}_\a {}^\g\e_\g A^i_\n\   +\mbox{c.c.}   \Rc  \RP
\ea\eqn{delta-term}
Setting to zero the external fields $u$, $v$, etc., we get the massive
action  
\eq\ba{l}
\left.\S \right\vert_{u=v=U_{ab}=V_{ab}=0}
 = \nmSI + \intx \Lp  
 \dfrac{1}{2g^2} \ml \Ml + \half \mp{ab}M^{\rm SUSY}_{ab}   
  -\dfrac{1}{4} \mffb{ab}\Mffb{ab}
\es\phantom{\left.\S \right\vert}%{u=v=U_{ab}=V_{ab}=0}=\nmSI + \intx}
   + \dfrac{1}{4} \mff{ab} \Mff{ab} +\l_{(3)abc}\f_a\f_b\f_c \Rp 
   +\ \mbox{c.c.}\ + O(\e) \ ,
\ea\eqn{massive-action}
 where the masses and the trilinear coupling are given by
\eq\ba{l}
\ml=2\k g^2\ ,\es
\mp{ab} = 2K_{ab} \ ,\es
\mffb{ab} = -4\k\kb \para{3}{}_{ab} +\k\kb\para{2}{}_{ac}\bpara{2}{}_{bc}
            +4K_{ac}\bar K_{bc}\ ,\es
\mff{ab} = 
 4\k K_{cd}\para{5}{}_{(cd)(ab)} + 2\k \lp K_{ca}\para{2}{}_{bc}
                             + K_{cb}\para{2}{}_{ac} \rp  \ ,\es
\l_{(3)(abc)} = \k\para{1}{}_{(abc)}\ .
\ea\eqn{masses}
We see that,  besides the \susyc\ mass\footnote{To which
belongs actually the term in 
$K_{ac}\bar K_{bc}$, which corrects for the fact that the \susyc\ mass term
\equ{susy-mass} was defined by the supersymmetry transformation
rules of the massless theory.}, 
all the mass terms belonging to the
GG class are present. A coupling trilinear in the scalar
matter field $\f$ appears, too, thus completing the list of the GG
``soft'' breakings.

%#############################################################
\section{Renormalization}\label{renormalisation}
The quantization of the theory runs according to the usual procedure of
studying its stability under radiative corrections and the determination
of the possible anomalies~\cite{piguet-sorella}. 
This analysis has been carried out
in~\cite{I} for the massless case, and it basically remains unaltered in
the massive case thanks to the particular way of introducing masses,
which does not change the cohomological sector of the theory. 
%#############################################################
\subsection{Stability of the Classical Theory}\label{stabilite}
The stability amounts to show that all possible perturbations of the
classical action \equ{complete-action} can be absorbed  by a
redefinition of its parameters. 
  These perturbations give all possible invariant counterterms which
can be freely added to the action at the quantum level, at each order in
$\hbar$.

By imposing that the perturbed action 
$\S+\z\S_{\rm c}$  satisfies
all the constraints defining the theory, at first order in 
the infinitesimal parameter $\z$ we find 
\eq
\S_{\rm c} = \S_{\rm nontriv} + \S_{\rm triv},
\eqn{pert-action}
where the  nontrivial counterterm can be written as
\eq
\S_{\rm nontriv}= Z_g\dpad{\S}{g} + Z_{(abc)}\dpad{\S}{\l_{(abc)}}
+ \bar{Z}_{(abc)}\dpad{\S}{\lb_{(abc)}}\ ,
\eqn{phys-ct}
and the trivial one as
\eqap
&&\S_{\rm triv} =  \BB_\S\lp\hat\S_1 + \hat\S_2\rp\ ,
\label{ct-hat}\\[4mm]
&&\hat\S_1 = \intx\LP
      Z^A (A^{*i\m}+\partial_\m\cb^i)A^i_\m + Z^\l \l^{*}{}^{i\a}\l^i_\a
      + {\bar {Z^\l}} \lb^{*}{}^i_\db\lb^{i\db}
\nonumber\es&&\phantom{\hat\S_1}
+ \Lp Z^\phi_{ab} \f^{*}_a\f_b + Z^\p_{ab} \p^{*}{}^\a_a\p_{b\a}
      + Z^u u\dfud{\S}{v} +Z^U_{(ab)(cd)} U_{ab}\dfud{\S}{V_{cd}}
      - \mbox{ c.c.} \Rp              \RP  ,
\label{ct-hat1}\es
&&\hat\S_2 = \intx\Lp 
      \ul \lp %Z_\ml\Ml + 
               \zpara{1}_{(abc)}\f_a\f_b\f_c
              +\zpara{2}_{ab} \f_a\e^\a\p^*{}_{b\a} 
              + \zpara{3}_{ab} \vbhat \Mffb{ab} 
              +\zpara{4}_{ab}\ub\e^\a \p_{a\a}\fb_b   \right. \nonumber\es
&&\phantom{\hat\S_1}   \left.
     %+ Z_{\mp{}} U_{ab} M^{\rm SUSY}_{ab} 
     + \zpara{5}_{(ab)(cd)}\Vhat_{ab}\Mff{cd}
     +\zpara{7}\e^\a\s^\m_{\a\db}\lb^{*i\db}A^i_\m  \rp
     +\zpara{6}_{(ab)c}^i\pa^\m U_{ab}\fb_c A^i_\m
%\nonumber\es&&\phantom{\hat\S_1}
     \  -\ \mbox{c.c.} \Rp\ .\label{ct-hat2}
\eea
As it can be seen from \equ{phys-ct}, the structure of the counterterm 
$\S_{\rm nontriv}$ is the same as in the massless case. 
It corresponds to
renormalizations of the gauge and Yukawa coupling constants. On the
other hand, the trivial cocycle $\S_{\rm triv}$ consists of two 
parts. 
The first one, $\BB_\S\hat\S_1$, is the same as in the massless 
case, and
contains the renormalizations of the fields, i.e. it corresponds
to anomalous dimensions. The second contribution is new.
 It yields the
renormalization of the additional parameters $\para{}$ 
appearing in the most
general solution \equ{complete-action}  of the \slav\ identity, hence in
particular of the masses through the definitions \equ{masses} --
except for the renormalization of $\ml$ and $\mp{ab}$, 
which is already given by that of
the doublets of external fields in \equ{ct-hat1}. However an
independent renormalization of the latter masses is
given by a possible redefinition
$\k\to\k'$ and $K_{ab}\to K'_{ab}$ of the shifts in the 
expressions \equ{ext-fields-doublets} which appear explicitly in the
\slav\ operator.

%#############################################################
\subsection{The Symmetries at the Quantum Level}\label{sym-quantiques}
For the quantum extension of the symmetries of the theory, one
can easily prove that the introduction of the shifted external
doublets does not affect the validity to all orders of perturbation
theory of the gauge condition \equ{gaugecond}, of the ghost equation 
\equ{ghostexpr} and of the global ghost equations \equ{globalgh}. In the
following we shall prove that it is possible to define a quantum vertex
functional $\G$ such that the condition 
of gauge invariance \equ{gauge-inv-mass}, the condition of
supersymmetry \equ{susy-cond} and the \slav\ identity \equ{class-slavnov}
hold to all orders:
\eq
X\G^{(0)} = \intx\dfud{\G^{(0)}}{u} =0 \ ,
\eqn{quantum-gauge-inv-mass}
\eq
Y_{ab}\G = \intx\dfud{\G}{U_{ab}} = 0\ .
\eqn{susy-cond-quantum}
\eq
\SS(\G) = 0\ ,
\eqn{quantum-slavnov}
where the superscript  ${}^{(0)}$
means that the corresponding quantity must be taken
at vanishing global ghosts.

Let us start with the first  identity \equ{quantum-gauge-inv-mass}.
According to the quantum action principle (QAP)~\cite{qap}, 
the quantum extension
of the corresponding classical identity is 
\eq
X\G^{(0)} = \D^{(0)}_u + O(\h\D^{(0)}_u)\ , 
\eqn{qap-gauge-inv-cond}
where the right-hand side is a breaking whose lowest nonvanishing order
in $\hbar$ is an
integrated local functional $\D^{(0)}_u$. 
Due to the anticommutativity of the operator
$\d/\d u$, $\D^{(0)}_u$ must
satisfy the consistency condition
\eq
X \D^{(0)}_u  = 0\ .
\eqn{consist-g-inv-cond}
 One easily checks that the most general solution of the
latter equation has the form
\eq
\D^{(0)}_u = X\D_X^{(0)}\ ,
\eqn{sol-gic}
where $\D_X^{(0)}$ is an integrated local functional with zero ghost
number and $R$-weight, of  dimension $\le4$. 
This implies that, after absorption of 
$\D_X^{(0)}$ as a counterterm, the functional $\G$ will obey 
\equ{quantum-gauge-inv-mass} at each order in $\hbar$.

Coming to the identity \equ{susy-cond-quantum}, we get from the QAP:
\eq
Y_{ab}\G  = \D_{ab} + O(\h\D_{ab}) \ ,
\eqn{qap-susy-cond}
where $\D_{ab}$ is an integrated local functional.
  Due to the anticommutativity of the operators 
$\d/\d U_{ab}$, $\D_{ab}$ must
obey the consistency condition
\eq
Y_{ab}\D_{cd} + Y_{cd}\D_{ab}  = 0\ .
\eqn{consist-susy-cond}
Because of the condition \equ{quantum-gauge-inv-mass} it must
satisfy the additional constraint
\eq
X\D^{(0)}_{ab}=0\ .
\eqn{cross-consist}
It is straightforward to find that the 
most general solution of the two latter conditions reads
\eq
\D_{ab} = Y_{ab} \D_Y\ ,
\eqn{sol-c-susy}
 where $\D_Y$ has the right dimension and quantum numbers.
Moreover, it may be chosen such as to obey the condition
\eq
X\D_Y^{(0)} = 0\ .
\eqn{L-delta0}
This means that \equ{susy-cond-quantum} can be made to hold, too,
without spoiling \equ{quantum-gauge-inv-mass}.

Finally, the \slav\ identity acquires a quantum breaking:
\eq
\SS(\G) = \D_{\rm ST} + O(\h\D_{\rm ST})\ ,
\eqn{qap-slavnov}
$\D_{\rm ST}$  being an integrated local functional of ghost number 1,
$R$-weight 0 and dimension bounded by 4,
constrained by the consistency conditions
\eq\ba{l}
\BB_\S \dslav = 0\ ,\es
X\D_{\rm ST}^{(0)} = 0 \ ,\es
Y_{ab}\dslav = 0 \ .
\ea\eqn{cons-slavnov}
The first condition is the Wess-Zumino consistency condition, which is
solved by
\eq
\dslav = \dsum{i} r_i\AA_i + \BB_\S\hat\D\ ,
\eqn{sol-cons-slavnov}
where the  ``anomalous'' terms
$\AA_i$ form a basis of the cohomology of the nilpotent
operator $\BB_\S$. With respect to the massless case~\cite{I}, no new
anomaly of this kind appears in the present massive case.
 In order to see this, let us introduce an expansion in the number
$n$ 
of doublet fields $u$, $\vhat$, $U_{ab}$ and $\Vhat_{ab}$,
i.e. such that each term of the expansion
is homogeneous of degree $n$ in these fields, and let us denote by $b_0$ the
term of $\BB_\S$ of degree~0:
\eq
b_0 = \nmBnmS +\intx\lp v'\dfud{}{u} + V'_{ab}\dfud{}{U_{ab}} \rp\ ,
\eqn{b0}
where we have set (c.f. \equ{BRS-u-v})
\eq
v' = \vhat - i\xi^\m\pa_\m u + \eta R_u u \ ,\quad
V'_{ab} = \Vhat_{ab} - i\xi^\m\pa_\m U_{ab} + \eta R_{U_{ab}}U_{ab} \ .
\eqn{v-prime}
 $b_0$ is nilpotent and 
does not change the degree. We have
\eq
b_0 u = v'\ ,\quad b_0 v' = 0\ ,\quad 
b_0 U_{ab} = V'_{ab}\ ,\quad b_0 V'_{ab} = 0\ ,
\eqn{b0-doublets}
This characterizes the pairs $(u,v')$ and  $(U_{ab},V'_{ab})$ as
``$b_0$-doublets''. It is known\footnote{See e.g. 
Section 5.2 of~\cite{piguet-sorella}.} that the
$b_0$-cohomology does not depend on such doublets, and also that the
cohomology of $\BB_\S$ is isomorphic to a subspace of the $b_0$-cohomology.

The only element $\AA_i$  in \equ{sol-cons-slavnov} 
is thus the supersymmetric
extension of the usual gauge anomaly~\cite{I},  whose coefficient will be
assumed to vanish through
a suitable choice of the matter field representation.
Therefore the \slav\ identity can be established by absorption of 
$\hat\D$ as a counterterm. 
 It can be checked that $\hat\D\ $
can be chosen in such a way that
it has the correct quantum numbers (ghost number 0, 
$R$-weight 0) and 
satisfies the conditions
\eq
X{\hat \D}^{(0)} = 0  \ ,\quad
Y_{ab}\hat\D = 0 \ \ ,
\eqn{cons-hat}
which are necessary for absorbing $\hat\D\ $ as a counterterm
without spoiling the two identities
\equ{quantum-gauge-inv-mass} and 
\equ{susy-cond-quantum}.

The absence of anomalies, together with the stability of the model, 
ends the proof of the renormalizability of the model.

%#############################################################
\section{Callan-Symanzik Equation}\label{eq-de-callan-symanzik}
  From now on we shall change from the parametrization
\eq
\lac g,\, \l_{abc},\, \k,\, K_{ab},\, \paras{2}{}_{ab},\, 
  \para{3}{}_{ab},\,  \para{1}{}_{(abc)},\, \paraa{2}{}_{ab},\, 
\para{4}{}_{ab},\, \para{5}{}_{(ab)(cd)},\, \para{6}{}_{(ab)c}^i,\,\para{7} \rac
\eqn{old-param}
to the new parametrization
\eq
%\lac p_I \rac :=
\lac g,\, \l_{abc},\, \ml,\, \mp{ab},\, \mff{ab},\, 
  \mffb{ab},\, \l_{(3)(abc)},\, \paraa{2}{}_{ab},\, 
  \para{4}{}_{ab},\, \para{5}{}_{(ab)(cd)},\, 
  \para{6}{}_{(ab)c}^i,\,\para{7} \rac
\eqn{new-param}
with the change of variables given by \equ{masses}. 
Taking into account the fact 
that $\mff{}$ is symmetric, we have
separated  the matrix $\para{2}$ in its symmetric and 
antisymmetric parts: 
\eq
\para{2}{}_{ab} = \paras{2}{}_{(ab)} + \paraa{2}{}_{[ab]}  \ .
\eqn{sym-antisym}
The parameters \equ{new-param} constitute 
the set of  physical parameters of the theory. 
The remaining  parameters correspond to the unphysical
field amplitude renormalizations compatible with the symmetries
 shown in \equ{ct-hat1}. 

All the masses in particular being defined by 
normalization conditions, the
shifts $\k$ and $K_{ab}$ are no longer independent parameters. They are
are thus renormalized in $\k'=\k+(\hbar)$ and $K'_{ab}=K_{ab}+O(\hbar)$.
The latter quantities replace the shifts  $\k$ and $K_{ab}$  
defined in \equ{ext-fields-doublets} and appearing explicitly in the
\slav\ operator (see \equ{ext-slavnov-op}).

The Callan-Symanzik equation describes the behaviour of the quantum
theory under  scale transformations. It may be obtained as 
follows~\cite{piguet-sorella}.
First, apply to the effective action the generator of scale
transformations 
\eq
D:= \m\dpad{}{\m} + \dsum{m} m\dpad{}{m}\ ,
\eqn{scale-op}
 where $\m$ is the mass scale at which  the
normalization conditions defining the parameters of the 
quantum theory are taken, and where
the summation over  $m$ extends over all masses and dimensionful coupling
constants. This defines, through the QAP, 
an insertion of dimension $\le4$. Second, expand the insertion $D\G$ 
in a suitable basis with the same dimensional constraints. 

In order to keep with the symmetries of the problem it is convenient to
work  with invariant insertions. 
An invariant insertion is by definition an insertion $\D$ obeying 
the conditions
\eq\ba{l}
\BB_\G (\D\cdot\G) = 0\ ,\es
\FF^i (\D\cdot\G) = 0\ ,\quad \pa_{\xi^\m}(\D\cdot\G) = 0\ ,\quad
  \pa_\eta(\D\cdot\G) = 0\ ,\es
X(\D\cdot\G)^{(0)} = 0\ ,\quad Y_{ab}(\D\cdot\G) = 0\ .
\ea\eqn{inv-insertion}
Such an invariant insertion is generated through the QAP
by the application to the effective action of a symmetric
operator $\na$: 
\eq
\na\G = \D\cdot\G \ ,
\eqn{qap-nabla}
 i.e. of an operator $\na$ which obeys the conditions
\eq\ba{l}
\na \SS(F) = \BB_F \na F\ ,\quad \forall F\ ,\es
\lc \na,\FF^i\rc = 0\ ,\quad 
\lc \na,\pa{}_{\xi^\m}\rc = 0\ ,\quad
\lc \na,\pa{}_{\eta}\rc = 0\ ,\es
\lc \na, X\rc^{(0)} = 0\ ,\quad \lc \na,Y_{ab}\rc = 0\quad \ .
\ea\eqn{sym-op}
and
\eq
\na\D^i_{\rm g}= 0\ ,\quad \na\D_\m^{\rm t}= 0\ ,\quad 
  \na\D_{\rm R}= 0\ ,
\eqn{sym-op'}
where $\D^i_{\rm g}$, $\D_\m^{\rm t}$ and $\D_{\rm R}$ are the
classical insertions appearing in the right-hand sides of the identities 
\equ{ghostexpr} and \equ{globalgh}.
The invariance of the insertion \equ{qap-nabla} then
follows from the effective action $\G$ fulfilling 
the quantum extensions of  the identities
\equ{class-slavnov}, \equ{ghostexpr}, \equ{globalgh}
\equ{gauge-inv-mass} and \equ{susy-cond}.

Let us look for a basis of invariant insertions.
The renormalized shifts $\k'$ and $K'_{ab}$ depending on the
parameters of the list \equ{new-param}, the partial derivatives of the
latter are made symmetric by redefining them according to 
\eq\ba{l}
\na_{p_I}:= \pad{}{p_I} - \intx\lp\pad{\k'}{p_I}\dfud{}{v}+
    \pad{\kb'}{p_I}\dfud{}{\bar v}
 + \pad{K'_{ab}}{p_I}\dfud{}{V_{ab}}
 + \pad{{\bar K}'_{ab}}{p_I}\dfud{}{\bar V_{ab}} \rp\ ,\\[4mm]
\lac p_I \rac :=
\lac g,\, \l_{abc},\, \mff{ab},\, 
  \mffb{ab},\, \l_{(3)(abc)},\, \paraa{2}{}_{ab},\, 
  \para{4}{}_{ab},\, \para{5}{}_{(ab)(cd)},\, 
  \para{6}{}_{(ab)c}^i,\,\para{7} \rac\ .
\ea\eqn{na-i}
 The mass parameters $\ml$ and $\mp{ab}$ are not included in the
latter list. Indeed, in the classical approximation they are identical to
the shifts $\k$ and $K_{ab}$ (c.f. \equ{masses}), 
and thus the corresponding $\na$
operators, giving zero when applied to the classical action, 
will not generate independent symmetric insertions.

 Further symmetric operators are the generators of the field
amplitude renormalizations compatible with the 
symmetries (c.f. \equ{ct-hat1}):
\eq\ba{ll}
\NN^A: = N^A -N^{A^*}-N^b-N^\cb\ ,\quad
&\NN^\l:= N^\l -N^{\l^*}    \quad\mbox{and c.c.}\ ,\es
\NN^\f_{ab} := \lp N^\f - N^{\f^*}\rp_{ab} 
   \quad\mbox{and c.c.}\ ,\quad
&\NN^\p_{ab} := \lp N^\p -N^{\p^*}\rp_{ab}\quad\mbox{and c.c.}\ ,\es
\NN^u :=  N^u + N^\vhat   \ ,\quad
&\NN^U_{(ab)(cd)} := \lp  N^U + N^\Vhat \rp_{(ab)(cd)}
      \quad\mbox{and c.c.}\ ,
\ea\eqn{field-ren}
where
\eq\ba{l}
N^\vf = \intx \vf\dfud{}{\vf}\ ,\quad \vf =
 A,\l,\lb,A^*,\l^*,\lb^*,b,\cb,u,\vhat,\bar u,\vbhat \ ,\es
N^\vf_{ab} = \intx \vf_a\dfud{}{\vf_b}\ ,\quad
  N^{\vf^*}_{ab} = \intx \vf^*_b\dfud{}{\vf^*_a}\ ,\quad
 \vf=\f,\p,\fb,\pb\ ,\es
N^\vf_{(ab)(cd)} = \intx \vf_{ab}\dfud{}{\vf_{cd}}\ ,\quad
  \vf = U,\Vhat,\bar U,\Vbhat\ .
\ea\eqn{count-op}
The operators $\NN$ are symmetric according to the definition 
\equ{sym-op} and \equ{sym-op'}, 
except $\NN^u$ and $\NN^U_{(ab)(cd)}$ for which
\eq
\lc \NN^u, X\rc = -X\ ,\quad 
\lc \NN^U_{(ab)(cd)},Y_{ef}\rc = -\d_{(ab)_(ef)} Y_{cd}\ .
\eqn{sym-lu-ku}
 where
\[
\d_{(ab)(cd)} = \lac\matrix{1\quad&\mbox{if}\quad ab = cd\mbox{ or } dc\cr
                            0\quad&\mbox{otherwise}\ .}\right.
\]
But this does not prevent them, together with the other ones,
to define symmetric insertions through \equ{qap-nabla}.

The set
\eq\ba{l}
\phantom{{\rm with}\quad} \lac \na_{p_I} ,\, \NN_K \rac\ ,\es
{\rm with}\quad           \lac \NN_K\rac = 
\lac \NN^A,\, \NN^\l,\, \NN^\f_{ab},\, \NN^\p_{ab},\, 
     \NN^u,\, \NN^U_{(ab)(cd)} \rac\ .
\ea\eqn{basis-sym-op}
forms a basis for the symmetric operators of the theory.
Their application to the effective action yields a basis
for the invariant insertions of dimension $\le4$. 

 The set of invariant insertions thus constructed
represents the  quantum extension
of the classical counterterms \equ{pert-action}.

On the other hand the symmetrized form 
\eq
\na_D:=D - \intx\lp \k'\dfud{}{v} + K'_{ab}\dfud{}{V_{ab}}
  +\mbox{ c.c.} \rp
\eqn{sym-scale-op}
of the scale operator \equ{scale-op}, where we have used the dimensional
analysis identities 
\eq
D\k'=\k'\ ,\quad DK'_{ab}=K'_{ab}\ ,
\eqn{dim-anal}
gives rise to an invariant insertion which may be expanded in the 
basis \equ{basis-sym-op}, yielding the Callan-Symanzik equation:
\eq
\lp \na_D + \dsum{I}\b_I\na_{p_I} - \dsum{K}\g_K\NN_K\rp \G = 0\ .
\eqn{CS}
The latter can be rewritten in a more explicit form 
 %(we set $u=v=U_{ab}=V_{ab}=0$)
\eq\ba{l}
\CC\G := \lp D + \dsum{I}\b_I\dpad{}{p_I} - 
\dsum{K}\g_K\NN^{\rm hom}_K\rp\G 
\es\phantom{R\G :} 
= \intx \lp \a_v\dfud{\G}{v} + \a_{\bar v}\dfud{\G}{\bar v}
   +   \a_{V_{ab}} \dfud{\G}{V_{ab}}
   +  \a_{\bar V_{ab}} \dfud{\G}{\bar V_{ab}} \rp \ ,
\ea\eqn{CS'}
where
\eq\ba{l}
\a_v = (\CC+\g_u)\k'\ ,\quad \a_{\bar v} = (\CC+\g_\ub)\kb'\ ,\es
\a_{V_{ab}} = \CC K'_{ab}+\g^U_{(ab)(cd)}K'_{cd} \ ,\quad 
                \a_{\bar V_{ab}} = \CC{\bar K}'_{ab}
    +\g^\Ub_{(ab)(cd)}{\bar K}'_{cd} \ , \es
\ea\eqn{rhs-CS}
and $\NN^{\rm hom}_K\ $ are the unshifted counting operators. 

\noindent{\bf Remark:\ }  The $\b_I$-terms for 
$p_I=\mff{ab},\,\mffb{ab}$ correspond to renormalizations of 
these mass
parameters. The renormalizations of the masses 
$\ml$ and $\mp{ab}$ are
expressed by the terms in the right-hand side of \equ{CS'}: they  
depend on the anomalous dimensions
of the external field doublets $(u,\vhat)$ and 
$(U_{ab},\Vhat_{ab})$.
This fact can already be seen by inspecting the structure of the
classical invariant counterterms \equ{ct-hat1}, \equ{ct-hat2}, 
the masses $\ml$ and $\mp{ab}$ 
being given by the shifts of $v$ and $V_{ab}$, respectively.

%#################################################################
\section{Conclusions}\label{conclusion} 

A realistic supersymmetric gauge field theory needs masses for the 
supersymmetric partners of the particles known up to now. In this paper, 
we introduced those mass terms which, according to~\cite{gir-gri}, induce 
supersymmetry breakings leading to less than quadratic divergences. In 
addition, we completed the theory by adding a mass term for the fermionic 
matter fields, preserving supersymmetry, besides classical gauge 
invariance. The main idea has been to introduce masses through a 
 constant shift of 
some external fields~\equ{BRS-u-v}. This 
allowed us to control the breakings due to the 
non-supersymmetric mass terms through the generalized  Slavnov-Taylor 
identity~\equ{quantum-slavnov},  collecting all the symmetries of the 
theory.
The issue of quantum gauge invariance of the mass terms 
has been characterized by the 
functional relation~\equ{quantum-gauge-inv-mass}.  
The quantum extension of the model has been 
performed according to the lines dictated by the Quantum Action 
Principle~\cite{qap} (see~\equ{quant-action-pr}),  both in 
the ultraviolet and in the infrared regions, 
this latter analysis being mandatory because of the presence 
 of both massive and massless fields. 
Moreover, the behaviour of the quantum 
theory under scale transformations has been characterized by 
the   Callan-Symanzik equation~\equ{CS}. 
The massive quantum theory constructed 
in this paper 
will be the starting point for the analysis 
of the  nonrenormalization properties of a realistic supersymmetric 
theory~\cite{futur}. 
%###############################################################
\appendix
\renewcommand{\theequation}{\Alph{section}.\arabic{equation}}
\renewcommand{\thesection}{\Alph{section}}
%++++++++++++++++++++++++++++++++++++++++++++++++++++++++++++
\setcounter{equation}{0}
\setcounter{section}{1}

%##########################################################
\section*{Appendix:%\thesection .
\ Infrared Power Counting}
\subsection{Ultraviolet and Infrared Subtractions}\label{soust-IR-UV}
In order to assure the existence of the Green functions, subtractions
have to be made. The momentum space
subtraction scheme of Zimmermann and 
Lowenstein~\cite{zim,low} assures their existence as tempered
distributions, free from UV and IR singularities, provided
the UV dimensions $d$ and IR dimensions $r$ 
of the fields fulfill certain conditions spelled in~\cite{low} (see below).
The method consists in subtracting off
the first few terms of the Taylor
expansion of the integrand of a divergent integral, corresponding to a
Feynman graph or subgraph, in its external
momenta $p$  and a certain parameter $s$ around \mbox{$p=s=0$}. 
The parameter $s$ appears
through the introduction of an auxiliary mass \mbox{$M(s-1)$} in the
denominator of every massless propagator -- here the gauge field and
ghost propagators. These subtractions, performed  recursively on the
integrands of the graph and of all its divergent one-particle-irreducible
(1PI) subgraphs,  make the integral UV-finite
without introducing spurious IR singularities thanks to the subtraction
terms involving only massive propagators. 
At the end $s$ is set to its physical value 1.
The Taylor expansion of the integrand associated to a divergent 
1PI 
graph or
subgraph $\g$  ends at the order defined by the ``UV degree of
subtraction'' 
\eq
\d(\g) = 4-\dsum{E}{}d_E\ ,
\eqn{uv-degree}
where the sum is performed on all the external lines of $\g$ and $d_E$
is the UV dimension of the corresponding quantum or external field.

However IR post-subtractions have to be performed as well, 
in order that subtracted subgraphs,
such as self-energy subgraphs of the massless fields, 
be kept vanishing at zero momentum. The latter
subtractions are done according to a Taylor expansion around $p=0$,
$s=1$, terminating at the order $\r(\g)-1$, where $\r(\g)$ is the ``IR
degree of subtraction''
\eq
\r(\g) = 4-\dsum{E}{}r_E\ ,
\eqn{ir-degree}
$r_E$ being the IR dimension of the corresponding field.

The conditions on the UV and IR dimensions are the following.
\begin{enumerate}
\item[-] If the propagator $\D_{AB}(k,s)$ 
between two quantum fields $\vf_A$ and
$\vf_B$ behaves asymptotically as
\[\ba{l}
\D_{AB}(\l p,\l s) \sim \l^{d_{AB}}\ ,\quad \l\to\infty,\ s\to\infty\ ,\es
\D_{AB}(\l p,\l(s-1)) \sim \l^{r_{AB}}\ ,\quad \l\to0,\ s\to1\ ,\es
\ea\]
then
\eq
d_A+d_B \ge d_{AB} + 4\ ,\quad r_A+r_B \le r_{AB} + 4\ .
\eqn{dim-A-B}
\item[-] For any (quantum or external) field $\vf_A$,
\eq
r_A \ge d_A\ .
\eqn{r-ge-d} 
\item[-] Some of the external momenta entering the graph or subgraph along
external lines corresponding to the fields $\vf_A$ may be set to zero
if and only if
\eq\ba{l}
\r_A \le 0\ ,\qquad\mbox{with the exception:}\es
\r_A \le \dfrac{3}{2}\ ,
  \quad\mbox{if only one external momentum is set to zero.}
\ea\eqn{r-le-0,3/2}
\end{enumerate}

%###############################################################
\subsection{Quantum Action Principle}\label{principe-d'action}
Let us define the UV dimension $d$ and the IR dimension $r$ of 
the functional derivative with respect to a field $\vf$ of dimensions
$d_\vf$ and $r_\vf$ as
\eq
d\lp \dfud{}{\vf(x)}\rp = 4-d_\vf\ ,\quad
r\lp \dfud{}{\vf(x)}\rp = 4-r_\vf\ .
\eqn{dim-fonct-der}
More generally the UV dimension of a functional operator, e.g. the operator of
a Ward or \slav\ identity, is defined as the maximum of the UV dimensions
of the individual terms
occurring in it, whereas its IR dimension is given by the minimum of the 
individual IR dimensions.
The quantum action principle (QAP)~\cite{qap} states that applying a
functional operator $\FF(x)$  %at point $x$ 
to the vertex functional yields the generating
functional of the vertex functions with a local insertion $Q(x)$ of
bounded UV and IR dimensions.  Similarly  for the partial
derivative with respect to a parameter $p$ of the theory. More precisely
\eq\ba{l}
\FF(x)\G = Q(x)\cdot\G\ , \qquad d(Q)\le d(\FF)\ ,\quad r(Q)\ge r(\FF)\ ,\es
\dpad{\G}{p} = \intx R(x)\cdot\G\ , \qquad d(R)\le 4\ ,\quad r(R)\ge 4\ ,
\ea\eqn{quant-action-pr}
where $Q(x)$ and $R(x)$ are local field polynomials.

An insertion characterized by the bounds ``UV dimension $\le d$,
IR dimension $\ge r$'' is called an $N_d^r$ normal product.
One notes, from \equ{r-le-0,3/2}, that integrated insertions must
be $N_d^4$, but also that an  $N_d^r$ integrated insertion 
with $4>r\ge5/2$ 
is allowed if it occurs only once. Moreover the insertions corresponding
to the interaction terms of the classical action must all be $N_4^4$, and
such must be the case of all counterterms.

%################################################################
\subsection{Absence of IR Singularities  %in the Present Model
}\label{IR-convergence}
The UV and IR dimension assignments given in Tables \ref{table-dim} and 
\ref{table-dim-u-v} fulfill the conditions \equ{dim-A-B}, \equ{r-ge-d}
and the $N_4^4$ prescription for the interaction terms of the classical
action. 

However, IR problems, namely IR anomalies,
may be encountered if insertions with $r<4$ have to be
absorbed in order to assure the validity of the functional
identities which define the symmetries of the theory, according to the
procedure followed in Section \ref{renormalisation}. 
Let us review the various cases at hand.

\noindent {\bf Identities \equ{quantum-gauge-inv-mass} and
\equ{susy-cond-quantum}:} In both cases an inspection of the possible 
candidate
counterterms $\D^{(0)}_X$ (see \equ{sol-gic}) and $\D_Y$ (see
\equ{sol-c-susy}), respectively, shows that all of them are $N_4^4$.

\noindent{\bf \slav\ identity \equ{quantum-slavnov}:} The 
situation 
is
slightly more tricky since, due to the inhomogeneity of the \slav\
operator the breaking $\D_{\rm ST}$ in \equ{qap-slavnov} is
$N_4^{5/2}$.
An inspection shows that all the possible counterterms 
$\hat\D$ (see \equ{sol-cons-slavnov}) have IR dimension 
$\ge 4$, but one:
\eq
{\hat\D}^3 = \intx\Vhat_{ab}\p^{*\a}_a\s^\m_{\a\da}\eb^\da A_\m\ ,
\eqn{ir-anom}
 which has IR dimension 3 due to the presence of a shifted external
field.
However, being linear in the quantum fields,
it can be absorbed as a counterterm since it will not appear in any 1PI
graph. 

In conclusion the theory presents no infrared anomaly, excepted the
harmless counterterm ${\hat\D}^3\ $.

\noindent{\bf Callan-Symanzik equation \equ{CS}:}
We finally come to the Callan-Symanzik equation, and doing so we
shall explain the  strange prescription for the IR 
dimension of
the external field $\l^*$ (see Table~\ref{table-dim}). 
Although the
Callan-Symanzik equation  does not belong to the defining properties of
the theory, its validity belongs to the current ``renormalization group''
understanding of  a 
renormalizable theory,  according to which any change of the energy-momentum
scale is compensable by a finite renormalization of the parameters of
the theory.
 
We first observe that the application of the symmetric operator $\na_D$ 
\equ{sym-scale-op} to the effective action yields an $N_4^3$ insertion.
The same holds for the basis of symmetric operators \equ{basis-sym-op}.
The Callan-Symanzik equation we have derived in
Section~\ref{eq-de-callan-symanzik} is thus perfectly compatible with
the renormalization scheme based on the dimensional
prescriptions of Tables~\ref{table-dim} and \ref{table-dim-u-v}.

\noindent{\bf Remark:\ }  Corresponding to the $a priori$ 
more natural choice for 
the IR dimension 5/2 of the external
field $\l^*$, the term of coefficient $\para{7}$ is absent from
the classical action \equ{complete-action} since it contains a part with
IR dimension 3 due to the shift in $\BB_\S u=\vhat$, and for the same
reason it does not appear as a counterterm: this prescription for
the IR dimension of $\l^*$  amounts to normalize $\para{7}$ to the value
zero. But a quantum extension of this same term
appears in the expansion of $\na_D\G$ in 
a basis of invariant insertions. Since in the absence of the parameter
$\para{7}$ it cannot be written as a partial derivative of the action,
the renormalization group interpretation of the Callan-Symanzik equation
is lost. We therefore have assigned the value 7/2 to the IR
dimension of $\l^*$, thus allowing for the presence of the parameter 
$\para{7}$ on the same footing as all the other ones and solving this
apparent paradox.

%#############################################################

\end{document}